\documentclass[aps,prl,twocolumn,longbibliography]{revtex4-1}

\usepackage{natbib}
\usepackage{xspace}
\usepackage{float}
\usepackage[sumlimits]{amsmath}
\usepackage{amsfonts,amssymb,amsthm}
\usepackage{graphicx}
\usepackage{physics}
\usepackage[margin=1in,footskip=0.25in]{geometry}
\usepackage{ulem} 

\usepackage{xr}
\usepackage{xcolor}

\newcommand{\TJWat}{IBM Quantum, IBM T.J. Watson Research Center, Yorktown Heights, NY 10598, USA}

\newcommand{\figfolder}[1]{}

\begin{document}

\title{Quantum crosstalk cancellation for fast entangling gates and improved multi-qubit performance}

\author{K. X. Wei}
\email{xkwei@ibm.com}
\author{E. Magesan}
\author{I. Lauer}
\author{S. Srinivasan}
\author{D. F. Bogorin}
\author{S. Carnevale}
\author{G. A. Keefe}
\author{Y. Kim}
\author{D. Klaus}
\author{W. Landers}
\author{N. Sundaresan}
\author{C. Wang}
\author{E. J. Zhang}
\author{M. Steffen}
\author{O. E. Dial}
\author{D. C. McKay}
\email{dcmckay@us.ibm.com}
\author{A. Kandala}
\email{akandala@us.ibm.com}


\affiliation{\TJWat}
\date{\today}

\begin{abstract}

Quantum computers built with superconducting artificial atoms already stretch the limits of their classical counterparts. While the lowest energy states of these artificial atoms serve as the qubit basis, the higher levels are responsible for both a host of attractive gate schemes as well as generating undesired interactions. In particular, when coupling these atoms to generate entanglement, the higher levels cause shifts in the computational levels that leads to unwanted $ZZ$ quantum crosstalk. Here, we present a novel technique to manipulate the energy levels and mitigate this crosstalk via a simultaneous AC Stark effect on coupled qubits. This breaks a fundamental deadlock between qubit-qubit coupling and crosstalk, leading to a 90ns CNOT with a gate error of (0.19 $\pm$ 0.02) $\%$ and the demonstration of a novel CZ gate with fixed-coupling single-junction transmon qubits. Furthermore, we show a definitive improvement in circuit performance with crosstalk cancellation over seven qubits, demonstrating the scalability of the technique. This work paves the way for superconducting hardware with faster gates and greatly improved multi-qubit circuit fidelities.

\end{abstract}

\maketitle

Existing quantum processors ~\cite{zhang2020,arute:2019} based on superconducting transmon qubits are pushing the limits of classical simulability. However, the realization of quantum advantage requires these processors to scale up in both size and operational fidelity. Reaching a suitable threshold on both counts would further enable quantum error correction and the realization of a fault tolerant quantum computer. These objectives require overcoming several technical challenges, notably, two-qubit gate fidelity, crosstalk, system stability and qubit coherence. One common architecture, based on fixed-frequency transmon qubits with fixed couplings, has a distinct advantage in terms of stability and coherence, but has limitations on gate speed and minimizing crosstalk due to always on interactions, and their relation to the exchange coupling strength, $J$. While a larger $J$ enables a faster entangling gate, the coupling leads to state dependent frequency shifts of neighboring coupled qubits, which is a source of quantum crosstalk that takes the form of a $ZZ$ interaction in the system Hamiltonian. This is formally seen from the standard cQED Hamiltonian for a pair of coupled transmons ($i = \{0,1\}$), modelled as Duffing oscillators,
\begin{eqnarray}
H_0/h & = & \sum_{i = \{0,1\}} \left(\nu_i \hat{a}_{i}^{\dagger}\hat{a}_{i} + \frac{\alpha_i}{2}\hat{a}_{i}^{\dagger}\hat{a}_{i} \left(\hat{a}_{i}^{\dagger}\hat{a}_{i}-1\right)\right)  \nonumber \\
&& +   J (\hat{a}_0^{\dagger}+\hat{a}_0)(\hat{a}_1^{\dagger}+\hat{a}_1),
\label{eqn:mainh_nodrive}
\end{eqnarray} 
with bare qubit frequencies $\nu_{i}$, bare anharmonicities $\alpha_{i}$ and coupling strength $J$. The coupling dresses the energy levels, and the crosstalk arising from state dependent frequency shifts is expressed as,
\begin{eqnarray}
\nu_{ZZ} & = & (\nu_{11}-\nu_{10})-(\nu_{01}-\nu_{00}). \label{eqn:zz_def}
\end{eqnarray} 
For fixed couplings, this is an always-on source of crosstalk, referred to as a static $ZZ$ interaction, with the following perturbative form,
\begin{eqnarray}
\nu_{ZZ,s}  & = &  -\frac{2 J^2 (\alpha_0+\alpha_1)}{(\alpha_1-\Delta_{0,1})(\alpha_0+\Delta_{0,1})},
\label{eqn:staticZZ2}
\end{eqnarray} 
where $\Delta_{0,1}$ represents the qubit-qubit detuning. This crosstalk has been seen to be an important limitation to multi-qubit circuit performance in tests of quantum volume~\cite{jurcevic:2020}, randomized benchmarking~\cite{mckay:2019}, and error correction codes~\cite{takita2016}, and may prevent device scaling ~\cite{berke:2020}.
Several hardware strategies have been employed to mitigate this crosstalk. The simplest approach, as seen from Eq.~(\ref{eqn:staticZZ2}), is to lower $J$, however, this comes at the expense of gate speed and lowers the overall gate fidelity due to finite qubit coherence. More involved strategies include the the introduction of tunable $J$ coupling ~\cite{chen2014,arute:2019,stehlik2021}; coupling different flavors of qubits with opposite signs of anharmonicity~\cite{zhao:2020,ku2020,xu2020} (see Eq.~(\ref{eqn:staticZZ2})); and the use of engineered multi-path coupling elements~\cite{mundada:2019,yan:2018,kandala2020,zhao2020,xu2020}. An alternative approach is a quantum control strategy to $ZZ$ cancellation via the AC Stark effect, using off-resonant radiation to selectively tune the energy levels, and modulate $ZZ$, as seen from Eq.~(\ref{eqn:zz_def}). This has been demonstrated with a single near-resonant, continuous wave (CW) drive in flux-tunable superconducting qubit architectures~\cite{noguchi2020,xiong2021}. However, this requires being close to a resonant transition outside the computational space, and is susceptible to charge noise in transmon qubits.

In this work we show that the $ZZ$ interaction for a pair of coupled transmon qubits can be tuned over several orders of magnitude by far-off resonant driving on {\it{both}} qubits. We develop an analytical model of the effect for transmons, building off previous theoretical work studying the case of coupled spins ~\cite{Li2008}. We then demonstrate that the effect, dubbed siZZle - Stark induced $ZZ$ by level excursions - can be employed for both static $ZZ$ cancellation as well as implementing $ZX$ and $ZZ$ entangling gates in all-transmon processors with simple direct capacitive coupling. The ability to cancel the static $ZZ$ interaction enables us to employ stronger qubit-qubit coupling, leading to a state-of-the-art cross-resonance gate with over a factor of 2 improvement in gate time from previous reports~\cite{kandala2020}. Furthermore, we demonstrate a novel high-fidelity CZ gate based on siZZle which adds to the toolkit of microwave-only~\cite{chow:2011,poletto:2012,chow:2013b,Paik2016,krinner:2020} two qubit gates. In contrast to previous approaches ~\cite{noguchi2020,xiong2021}, our approach with Stark tones on both qubits introduces an additional control parameter, the phase difference between the two tones, that is particularly useful for extending to larger devices. We demonstrate $ZZ$ cancellation on a line of 7 qubits combining siZZle with the hardware approach of multi-path couplers, and demonstrate improvements in the performance of Quantum Volume (QV) circuits~\cite{cross2019}.

\begin{figure*}[!ht]
\includegraphics[width = 1.5\columnwidth]{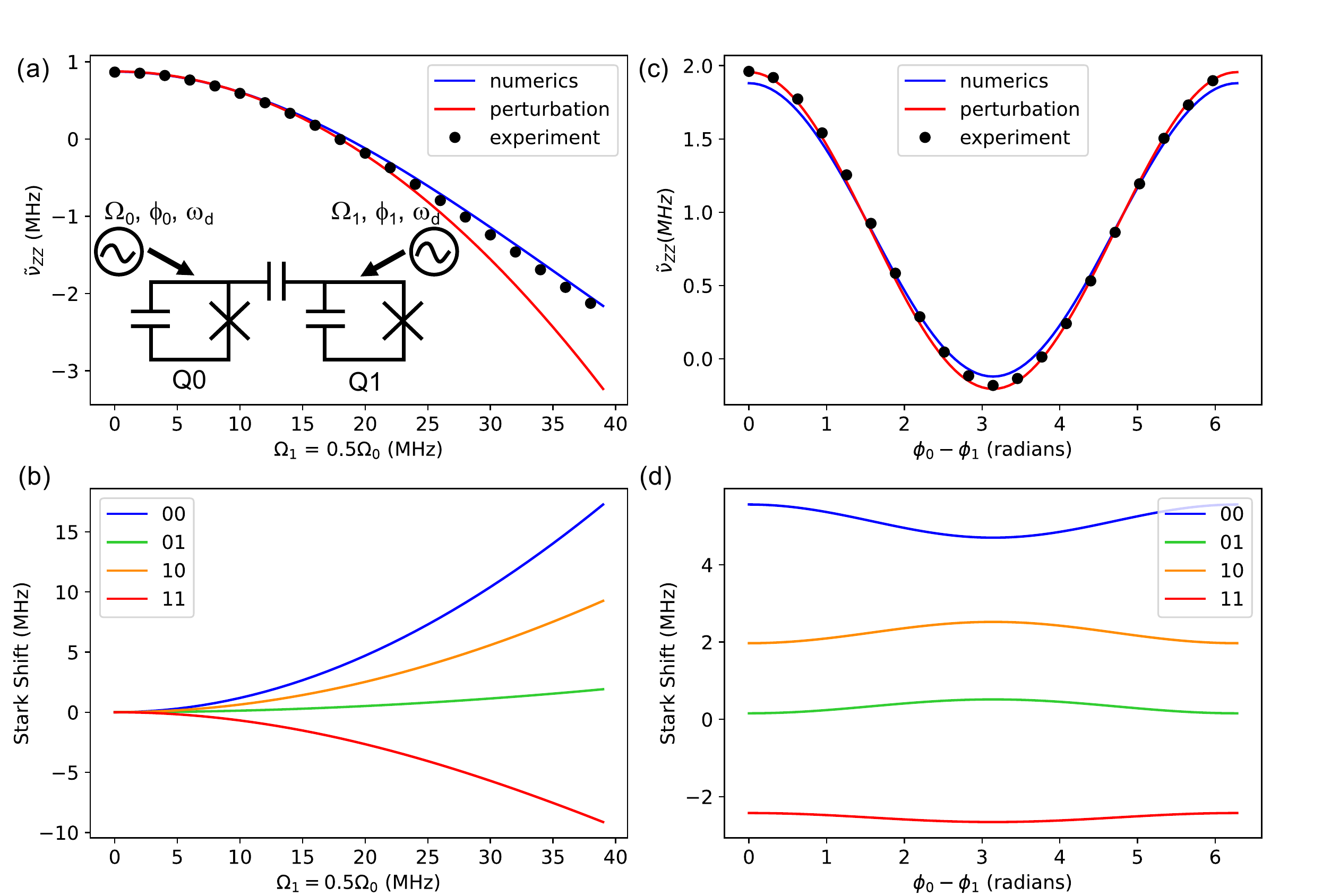}
\caption {{\bf {Physics of siZZle}} (a) Modulation of the $ZZ$ interaction strength $\tilde{\nu}_{ZZ}$ as the Rabi amplitude of the Stark tones is swept (ratio $\Omega_1/\Omega_0 = 0.5$) for fixed frequency $\nu_d = 5.075$ GHz and phase difference $\phi =\pi$. Experimental data (black circles) is compared to numerical (blue line) and perturbative (red line) calculations using the device parameters of Table 1. The inset shows a circuit representation of the primary two-qubit device discussed in this work. (b) The corresponding excursions of the computational levels, calculated numerically, to generate the $\tilde{\nu}_{ZZ}$ shown in (a).(c) Modulation of the $ZZ$ interaction strength $\tilde{\nu}_{ZZ}$ as the phase difference between the Stark tones is swept, for fixed frequency $\nu_d = 5.075$ GHz and and drive amplitudes $\Omega_1= 0.5\Omega_0 = 20$ MHz. Experimental data (black circles) is compared to numerical (blue line) and perturbative (red line) calculations using the device parameters of Table 1 (d) The corresponding excursions of the computational levels, calculated numerically, to generate the $\tilde{\nu}_{ZZ}$ shown in (c).
}
\label{fig:theory}
\end{figure*}

To describe the physics of siZZle, we consider the Hamiltonian of Eqn.~(\ref{eqn:mainh_nodrive}) and add off-resonant drives on both qubits,
\begin{eqnarray}
& & H_{\textrm{siZZle}}/h =  H_0/h +  \nonumber \\
& & \sum_{i = \{0,1\}} \Omega_{i}\cos{(2\pi \nu_{d}t+\phi_i)}(\hat{a}_i^{\dagger}+\hat{a}_i) , \label{eqn:mainh}
\end{eqnarray} 
with amplitudes $\Omega_i$, phases $\phi_i$, and a common frequency $\nu_d$ . The device schematic in Fig.~\ref{fig:theory}(a) depicts a simple direct capacitive coupling between the qubits that produces the Hamiltonian model of Eq.~\ref{eqn:mainh}. 
In the limit of ${\Omega_i}/{|\nu_d-\nu_i|} \ll 1$, we can write the dressed RWA Hamiltonian as,
\begin{equation}
H_{\textrm{eff}}/h=\tilde{\nu}_{ZI}{ZI}/4+\tilde{\nu}_{IZ}{IZ}/4+\tilde{\nu}_{ZZ}{ZZ}/4,
\label{eqn:effH}
\end{equation}
where the tilde notation refers to being in the doubly-dressed frame with respect to the exchange coupling and Stark tones. 
To second order in $\Omega_i$ and first order in $J$, the $ZZ$ coefficient is,
\begin{eqnarray}
\tilde{\nu}_{ZZ} & = &  \nu_{ZZ,s} + \nonumber \\
& & \frac{2J\alpha_0 \alpha_1\Omega_0 \Omega_1 \cos{(\phi_0-\phi_1)}}{\Delta_{0,d} \Delta_{1,d} (\Delta_{0,d}+\alpha_0)(\Delta_{1,d}+\alpha_1)}, \label{eqn:zz_sizz}
\label{eqn:ZZ}
\end{eqnarray} 
where the static term is given by Eqn.~(\ref{eqn:staticZZ2}). In the above equations, $\Delta_{i,j}=(\nu_i - \nu_j)$ denotes detunings where $i,j \in \{0,1,d\}$. The most significant contribution to the Stark shifts comes from the term associated with a single, isolated drive
\begin{eqnarray}
\tilde{\nu}_{ZI,\textrm{single}} = -\frac{\Omega_0 ^2\alpha_0}{\Delta_{0,d} (\Delta_{0,d}+\alpha_0)}, \label{eqn:zi_single}
\end{eqnarray} 
which will be of significance in later discussions for the impact of the Stark tones on qubit coherence. A formal derivation of these expressions is discussed in the Supplementary Information. 
Eq.~(\ref{eqn:zz_sizz}) reveals the various control knobs to manipulate the strength of the Stark induced $ZZ$ interaction: the amplitudes of the two tones, the drive-qubit detunings, the anharmonicities, and the phase differences between the two drive tones. 

\begin{figure*}[!ht]
\includegraphics[width = 1.9\columnwidth]{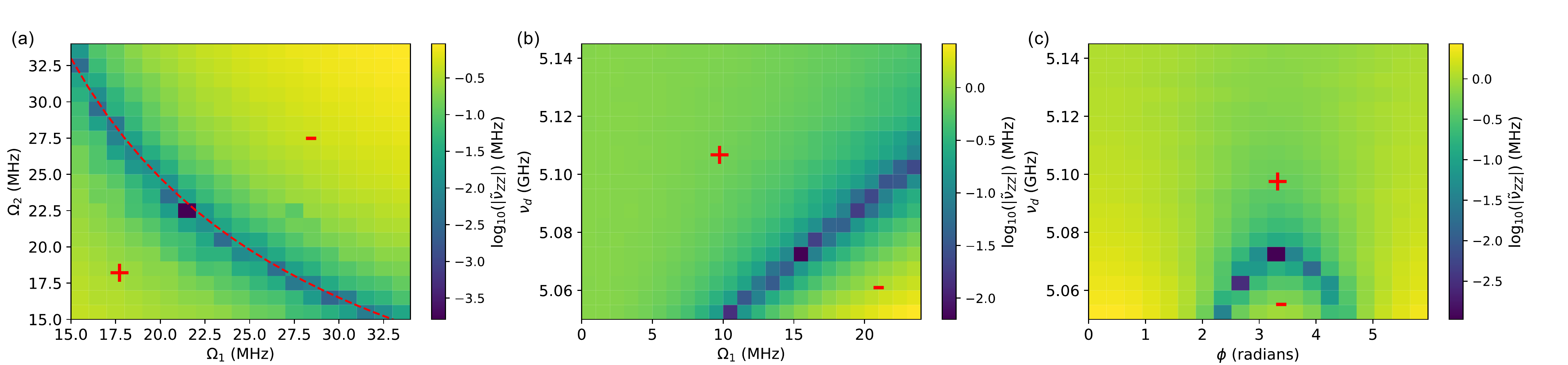}
\caption{ {\bf{Mapping the siZZle parameter space}} (a) Experimental sweep of $\tilde{\nu}_{ZZ}$ versus Stark amplitudes for fixed Stark frequency $\nu_d = $ 5.065 GHz and calibrated phase $\phi = \pi$. The red dotted line highlights the $\tilde{\nu}_{ZZ} \propto \Omega_0 \Omega_1$ dependence that is expected from the perturbative expression of \eqref{eqn:mainh}. (b) Experimental sweep of $\tilde{\nu}_{ZZ}$ versus Stark amplitude and Stark frequency for a fixed ratio of Stark amplitudes $\Omega_1/\Omega_0=0.4$ and calibrated phase $\phi = \pi$. (c) Experimental sweep of $\tilde{\nu}_{ZZ}$ versus the phase difference $\phi$ and Stark frequency for a fixed Stark amplitudes $\Omega_0= 37.5$ MHz, $\Omega_1=15$ MHz. The + and - symbols in the 3 sub-figures refer to the sign of $\tilde{\nu}_{ZZ}$.
}
\label{fig:2Dexpt}
\end{figure*}

\begin{figure*}[!ht]
\includegraphics[width = 1.9\columnwidth]{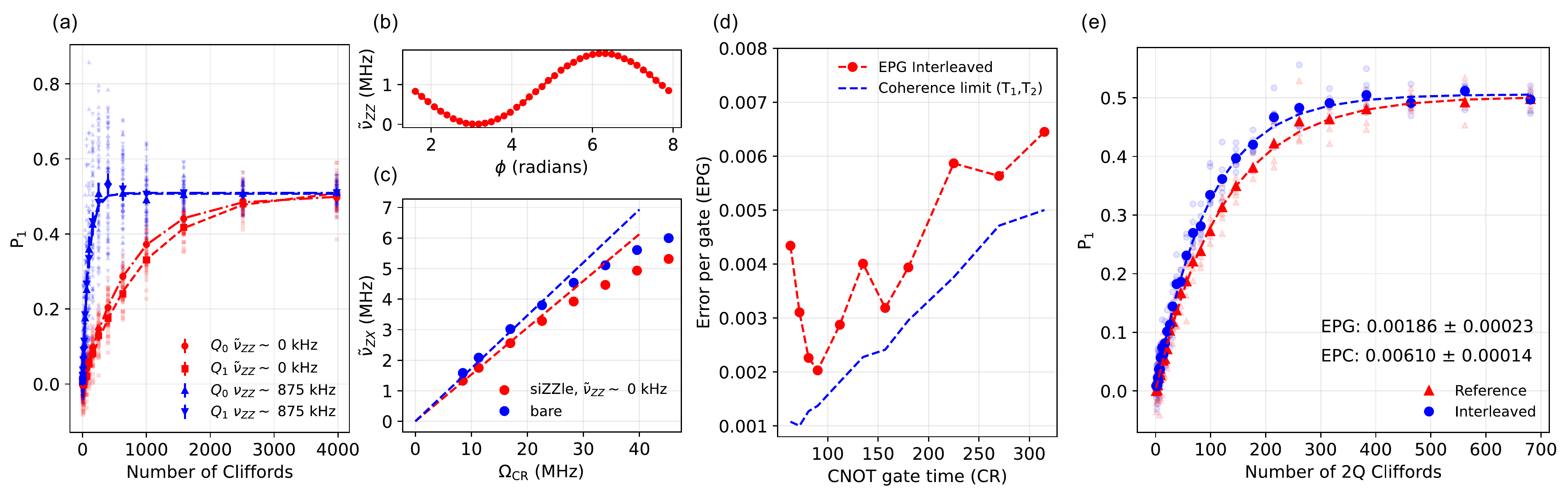}
\caption{{\bf{Fast cross-resonance with static ZZ cancellation}} (a) Simultaneous randomized benchmarking (RB) of 50~ns single qubit gates in the absence of static $ZZ$ cancellation (blue) leads to an average error per gate (EPG) of 6.6e-3. After static $ZZ$ cancellation with a pair of CW Stark tones at $\nu_d =$ 5.1 GHz, the EPG dramatically improves to 7.1e-4 (red). Bold symbols represent mean of the individual seeds (represented by light symbols), and dotted lines are exponential fits to the decay of the excited state probability $P_1$. (b) Phase calibration of the CW Stark tones to $\tilde{\nu}_{ZZ} \sim 0$ for $\Omega_0 = 59$ MHz and $\Omega_1 = 22$ MHz. (c) Strength of $ZX$ interaction $\tilde{\nu}_{ZX}$ versus cross-resonance drive amplitude $\Omega_{\textrm{CR}}$ with (red) and without (blue) static $ZZ$ cancellation. Here, Q1 is the control qubit and Q0 is the target qubit. Bold circles represent experimentally measured rates, using Hamiltonian tomography. Dotted lines are perturbative estimates, using Eq.~\ref{eqn:ZXsizzle}. (d) EPG measured by interleaved RB, for direct CNOT gates constructed from cross-resonance, after $ZZ$ cancellation, as a function of CNOT gate time. The blue dotted line represent the coherence limit to gate error from estimated using standard $T_1$ and $T_2$ measurements before every RB run. (e) Post-$ZZ$ cancellation interleaved RB of a 90 ns direct CNOT gate reveals a best EPG of 1.86e-3, with an upper bound on the EPG of 4.0e-3.}
\label{fig:ZZcan}
\end{figure*}

\begin{figure}[h!]
\includegraphics[width = 1.0\columnwidth]{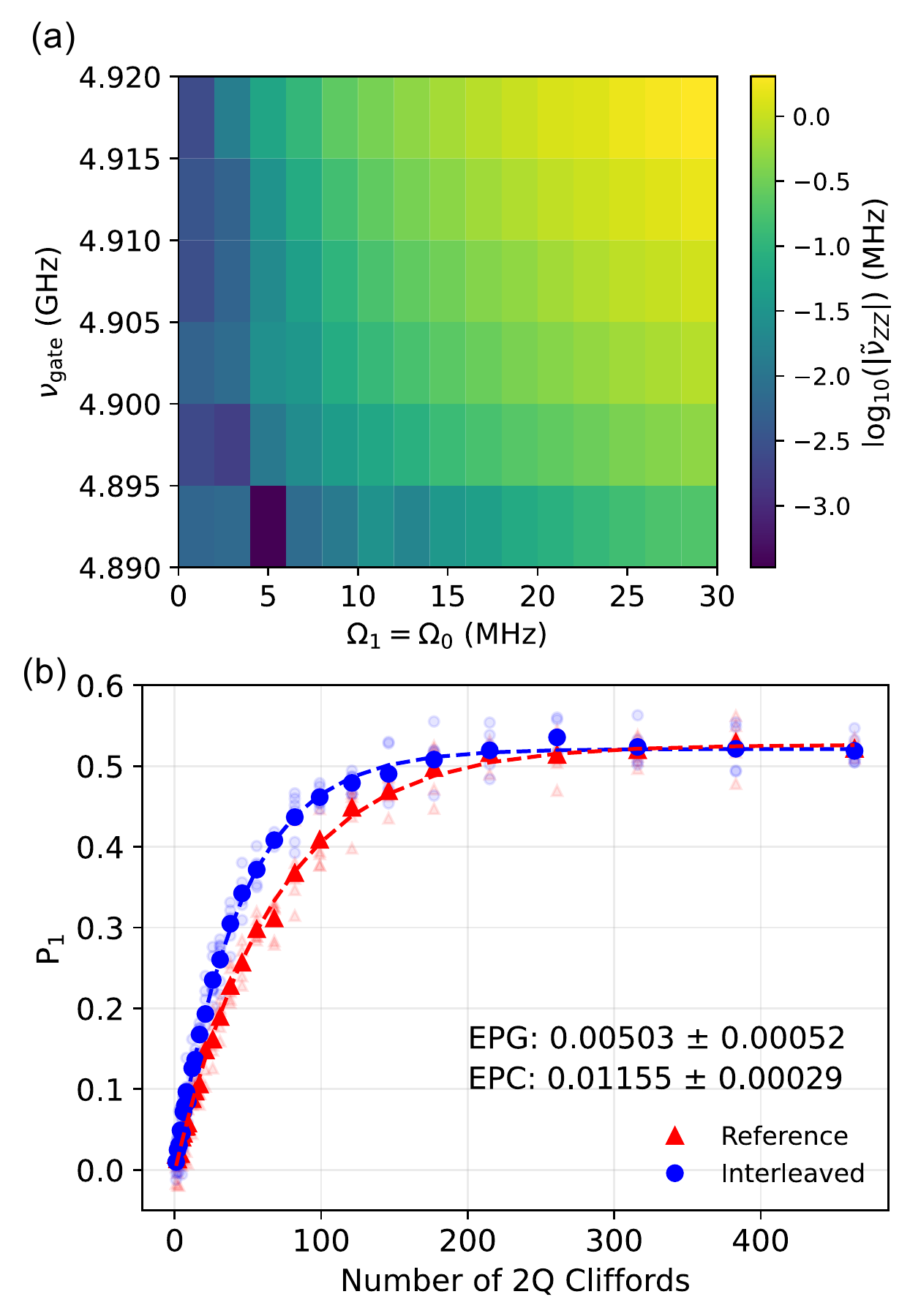}
\caption{ {\bf{All $ZZ$ SiZZle gate}} (a) Post ZZ cancellation 2D sweep of $\nu_{ZZ}$ with pulsed Stark frequency $\nu_{\textrm{gate}}$ and amplitude, with the ratio of the two amplitudes fixed to $\Omega_0 = \Omega_1$, and phase calibrated to maximum contrast. The CW tones to cancel $ZZ$ use the same parameters discussed in Fig \ref{fig:ZZcan}, with $\nu_d=5.1$~GHz. (b) Interleaved RB of a calibrated CZ gate based on siZZle reveals an error per gate of 5e-3, with an upper bound on that figure of 7.6e-3. }
\label{fig:Sizzlegate}
\end{figure}

Fig. \ref{fig:theory} reveals the physics of siZZle, employing the parameters of the primary two-qubit device studied in this work, device A. The parameters are given in Table \ref{table:deviceprop}. We perform numerical diagonalization of Eq. \eqref{eqn:mainh} after moving into the frame of the drive. Fig. \ref{fig:theory} depicts how the excursions of the computational levels leads to a modulation of $\tilde{\nu}_{ZZ}$, as the Stark tone amplitudes ((a), (b)) and phase difference ((c), (d)) are swept. We also see good agreement between the numerical calculations and the derived analytical expression of Eq.~\ref{eqn:ZZ} in the perturbative limit. Experimentally, we measure $\tilde{\nu}_{ZZ}$ by employing standard Ramsey sequences on Q0 while Q1 is in $|0\rangle$ or $|1\rangle$. The experimentally measured values show very good agreement with numerics in Fig.~\ref{fig:theory}(a), (c). A wider parameter space is experimentally mapped in the 2D sweeps of Fig.~\ref{fig:2Dexpt} and further depicts the physics of siZZle. Fig.~\ref{fig:2Dexpt}(a) maps $\tilde{\nu}_{ZZ}$ versus the Rabi amplitudes of the Stark tones on both qubits, and the region of $\tilde{\nu}_{ZZ} \sim 0$ kHz clearly highlights the $\tilde{\nu}_{ZZ}  \propto \Omega_0 \Omega_1 $ dependence expected from Eq.~(\ref{eqn:zz_sizz}). Fig.~\ref{fig:2Dexpt} (b) shows that modulation of $\tilde{\nu}_{ZZ}$ versus siZZle frequency and the Rabi amplitudes, and shows that sizeable $ZZ$ modulation can be obtained over a wide range of frequencies. As can be seen qualitatively from Eq.~(\ref{eqn:zz_sizz}), placing the Stark tone away from the qubit frequency can be compensated by increasing the drive amplitude, for the same $\tilde{\nu}_{ZZ}$.  Fig.~\ref{fig:2Dexpt} (c) demonstrates the sinusoidal phase dependence of $\tilde{\nu}_{ZZ}$, over a range of frequencies. The experimental data of Figures \ref{fig:theory} and \ref{fig:2Dexpt} reveal two interesting regimes of operation. At fairly modest drives, we observe see that we can cancel the $ZZ$ interaction to operate at  $\tilde{\nu}_{ZZ} \sim 0$. At stronger drive amplitudes, one can generate large $ZZ$ rates for two qubit entangling gates. These regimes of operation are discussed in Fig.~\ref{fig:ZZcan} and \ref{fig:Sizzlegate} and in the next two sections.

\begin{table}[h!]
\begin{tabular}{|c |c  |c  |c  |c  |c  |} 
 \hline
 $ $  & $\tilde \nu_0$  & $ \tilde \nu_1$   & $ \tilde \alpha_0$   &$ \tilde \alpha_1$    \\ [0.5ex] 
 \hline
 \textrm{No siZZle} & 4.960 & 5.016 & -0.283 & -0.287 \\ 
  \hline
  \textrm{siZZle} & 4.953 & 5.014 & -0.276 & -0.286 \\ 
  \hline
\end{tabular}


\caption{Qubit frequencies for device A depicted in Fig. \ref{fig:theory}(a) before and after $ZZ$ cancellation. All the numbers are in units of GHz. We note that these numbers represent the experimentally measured frequencies, dressed by the coupling $J=7.745$ MHz. }
\label{table:deviceprop}
\end{table}

In the first regime of operation, siZZle is used to cancel $ZZ$, which can be utilized to increase the speed of entangling gates, such as cross-resonance  (CR)~\cite{paraoanu:2006,chow:2011}, which are set by the coupling strength $J$. As discussed previously in Eq.~\ref{eqn:staticZZ2}, increasing $J$ typically leads to large values of static $ZZ$ crosstalk. 
Recent work~\cite{kandala2020} with multi-path couplers demonstrated a way to break the standard relationship between $J$ and $\nu_{ZZ,\textrm{static}}$ (operating at $J/\nu_{ZZ,\textrm{static}} \sim 130$), leading to state-of-the art CR gate fidelities. A drawback of the multi-path coupler approach is that $\nu_{ZZ,\textrm{static}}$ depends strongly on the qubit frequencies, so that attempting to achieve $\nu_{ZZ,\textrm{static}} \sim 0$ is non-trivial in fixed frequency architectures given current precision over qubit frequency allocation~\cite{zhang2020}. Our quantum control approach to $ZZ$ cancellation introduced here enables tuning to $\tilde{\nu}_{ZZ} \sim 0$ over a range of parameters since we have several degrees of freedom in our control space. Importantly, this allows for a decoupling of $J$ and $\tilde{\nu}_{ZZ}$ so that fast, high-fidelity entangling gates are possible with minimal static crosstalk in an architecture consisting of standard single path couplers and nominally fixed-frequency qubits. 

To test this, our device, described in Table~\ref{table:deviceprop} has a large coupling strength of $J \sim 7.745$ MHz, leading to a very large static $ZZ$ interaction of $\nu_{ZZ,\textrm{static}} = 875$ kHz. Without any further mitigation of $ZZ$, this prevents high-fidelity simultaneous single qubit operation due to strongly state-dependent qubit frequencies. This is seen in the decay and variance of simultaneous single qubit randomized benchmarking sequences shown in Fig. \ref{fig:ZZcan}(a) with an estimated average error per gate (EPG) of 6.6e-3. In order to mitigate this crosstalk, we add continuous wave (CW) Stark drives to cancel $ZZ$ and operate in a basis dressed by these off-resonant drives. The system Hamiltonian builds off Eq. \eqref{eqn:mainh} to now include additional drives for gate operation: 
\begin{eqnarray}
H/h & = & H_{\textrm{siZZle}}/h + \nonumber \\
&& \sum_{i=\{0,1\}} \Omega_{i,\textrm{gate}}(t)\cos{(2\pi\nu_{i,\textrm{gate}}t+\phi_i)}(\hat{a}_i^{\dagger}+\hat{a}_i) \nonumber
\label{eqn:mainh2}
\end{eqnarray} 
where $\Omega_{i,\textrm{gate}}(t)$ and $\nu_{i,\textrm{gate}}$ are the time-dependent amplitude and frequency of the single/two-qubit gate drive on qubit $i$ respectively.
 
The large choice of operating parameters for the $ZZ$ cancellation tones makes identifying an optimal set of working parameters a complex task. First, we limit leakage out of the computational subspace by placing the $ZZ$ cancellation tone above both qubits.  Next, we optimize the detuning of the cancellation tone. Smaller detuning reduces the drive amplitude required for $ZZ$ cancellation. There is a practical limit to the amount of amplitude that can delivered to the qubits before there is heating of system components. However, if the detuning is too small then the cancellation tone may start to interfere with the gate drive and time-dependent terms in the effective Hamiltonian in the frame of the drive can no longer be ignored.
For these reasons, we select $\nu_d =$ 5.1 GHz, for device A. The CW amplitudes are chosen to be sufficient to just approach $\tilde{\nu}_{ZZ} \sim 0$ after phase calibration (i.e at $\phi=\pi$), see Fig.~\ref{fig:ZZcan}(b). We estimate the CW amplitudes from the independent qubit Stark shifts to be $\Omega_0 = 59$ MHz and $\Omega_1 = 22$ MHz. After tuning to $\tilde{\nu}_{ZZ} \sim 0$, the single qubit gates are re-calibrated with the cancellation drives on. The new operating frequencies of the qubits are $\tilde{\nu_0} = 4.953$ GHz and $\tilde{\nu_1} = 5.014$ GHz, and so, the qubits have modest Stark shifts of -7.8 MHz and -1.7 MHz respectively. Reducing the $ZZ$ in this way results in remarkable improvements in simultaneous single qubit operation for 50 ns gates, with an estimated gate error of 7.1e-4 from randomized benchmarking, see Fig.~\ref{fig:ZZcan}(a). We note that there are several operating points for achieving $\nu_{ZZ} \sim 0$, but operating at stronger CW amplitudes with larger Stark shifts can to lead to additional dephasing. 

With $ZZ$ cancelled and single-qubit gates calibration, we now calibrate a two-qubit gate with cross-resonance. This entails additional drives on the control qubit (Q1) at the dressed target qubit (Q0) frequency. In Fig. \ref{fig:ZZcan}c, we measure the generated $ZX$ rates versus CR drive amplitude from tomography of the CR drive Hamiltonian, with and without $ZZ$ cancellation. The $ZX$ rate is modified due to the presence of the cancellation tones, however, as a consequence of the large $J$ coupling, one can access fairly large $ZX$ rates at modest CR drive amplitudes. A perturbative model for the $ZX$ rate is derived that includes the contribution from the cancellation tones. Assuming a CR tone on transmon 0 (for the experiment of this paper the CR tone is on transmon 1 so the labels will be swapped) we have,
\begin{align}
    \tilde{\nu}_{ZX} &\sim J\Omega_{0,\text{gate}}\left(A+B\Omega_0^2 + C\Omega_1^2\right),
\label{eqn:ZXsizzle}
\end{align}
where
\begin{align}
A &= -\frac{\delta}{\Delta_{0,1}(\delta+\Delta_{0,1})},
\end{align}
and $B$, $C$ are given in the supplement. We see that the $ZX$ rate has contributions that are quadratic in the cancellation tone amplitudes. The zero-point slope for the $ZX$ rate is modified by the Stark tones and when $\Omega_0=\Omega_1=0$ the usual first order expression for $\tilde{\nu}_{ZX}$ is obtained. Fig.~\ref{fig:ZZcan}c contains the $ZX$ rates with the Stark tones both off and on, and we see good agreement between the perturbative model and experiment at low CR amplitudes.

The large $J$ coupling is also of consequence for the reduced control qubit Stark shift, discussed previously in ~\cite{kandala2020}, and the resulting stability of unechoed {\it{direct}} CNOT gates constructed using CR. We construct and calibrate direct CNOT gates, similar to ~\cite{kandala2020}, and study the gate error obtained from interleaved RB as a function of CNOT gate time in Fig. \ref{fig:ZZcan}d. The calibration sequences and pulse shapes are detailed in the supplement. At the optimal gate time of 90 ns, we depict results from interleaved RB sequences in Fig. \ref{fig:ZZcan}e, that we use to estimate an error per gate (EPG) of 1.86e-3, with an error per Clifford (EPC) of 6.0e-3 from standard RB. Our decomposition has 1.5 CNOT gates per Clifford on average and this places an upper bound on the EPG of EPC/1.5 $\sim$ 4.0e-3. The ratio of EPG/EPC can be compared to analysis in ~\cite{epstein:2014} for confidence in the interleaved RB estimates. We also note that our gate errors fluctuate with changes in coherence and the defect environment~\cite{carroll2021} in the vicinity of the qubit frequencies. At the time of the displayed benchmarking, our measured coherence times for Q0(Q1) were $T_1 =$ 66 (66) $\mu$s and $T_2 =$ 49(84) $\mu$s.

\begin{figure}[!ht]
\includegraphics[width = 1.0\columnwidth]{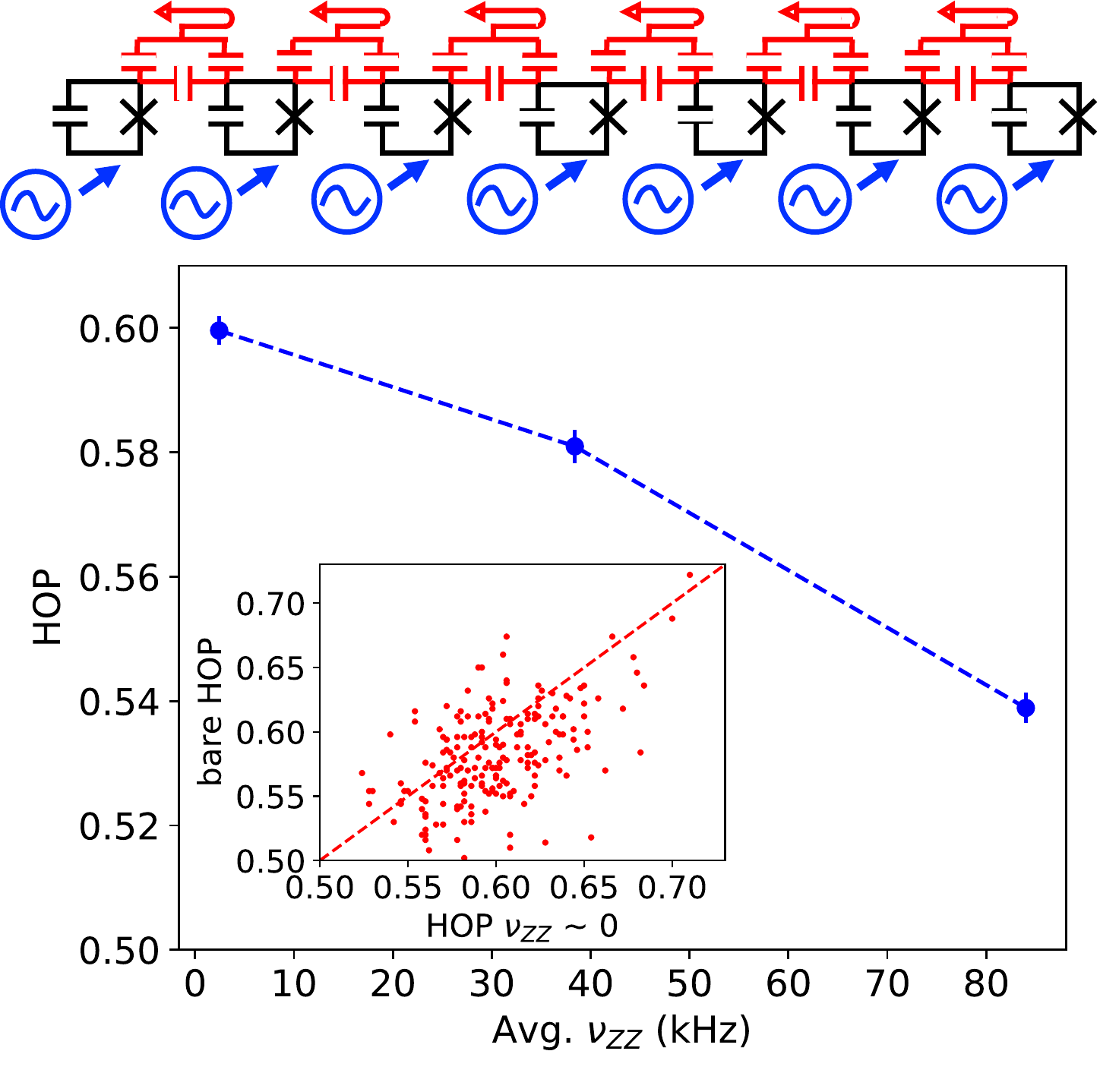}
\caption{ {\bf{Dependence of multi-qubit circuit fidelity on $ZZ$ interaction}} (Top) A device schematic of the line of 7 qubits, with a combination of hardware and control approaches to $ZZ$ modulation. The device employs multi-path couplers composed of a direct capacitive coupling and a $\lambda$/4 bus resonator. (Bottom) Average heavy output probability (HOP) for the same set of 200 random quantum volume (QV) circuits, at different levels of $\tilde{\nu}_{ZZ}$. Error bars represent standard error of the mean. The maximum and minimum $\tilde{\nu}_{ZZ}$ data points are tuned by setting the pair wise phase difference between the siZZle tones to $\phi \sim 0$ and $\phi \sim \pi$ respectively. The middle data point is measured in the absence of siZZle. (Inset) Scatter of individual circuit HOPs for the native (bare) device versus post-$ZZ$ cancellation. }
\label{fig:HOP}
\end{figure}

In the second regime of operation, siZZle can be used as a standalone method for performing a two-qubit gate due to the large $ZZ$ rates that can be generated as shown in Figs.~\ref{fig:theory} and \ref{fig:2Dexpt}. 
In order to mitigate the static $ZZ$, we continue to use CW tones at $\nu_d = $ 5.1 GHz, but, additionally pulse a second set of off-resonant tones at a different frequency $\nu_{\textrm{gate}}$ to generate large $\tilde{\nu}_{ZZ}$. This is shown in Figure \ref{fig:Sizzlegate}a, where we sweep the pulsed tone frequency and amplitudes ($\Omega_{0,\textrm{gate}} = \Omega_{1,\textrm{gate}}$) to generate $\tilde{\nu}_{ZZ}$ exceeding a few MHz. We note that $ZZ$ gate operation can also be achieved with a single frequency, using amplitude or phase modulation to switch between low and high $ZZ$ rates.
Once again, the operating parameter space is very large, and finding a parameter set that is optimized for gate fidelity, speed and leakage is a challenging task that is left for future study. Here, we provide a proof-of-concept example of siZZle gate operation at $\nu_{\textrm{gate}} = 4.9$ GHz, with maximum amplitudes $\Omega_{0,\textrm{gate}}, \Omega_{1,\textrm{gate}} \sim 26$ MHz. We calibrate the phase difference between the phase tones for maximum $\tilde{\nu}_{ZZ}$, and employ frame changes on the control and target qubits to construct a novel {\it {direct}} CZ gate of length 200 ns. Interleaved RB, shown in Fig. \ref{fig:Sizzlegate}b reveals a gate error of 5e-3, with an error per gate upper bound of 7.6e-3.

Finally, we study the impact of siZZle on multi-qubit circuit fidelity, using a line of 7 qubits from a 27 qubit device with a heavy-hex architecture~\cite{jurcevic:2020}, that we shall refer to as Device B. In order to reduce the impact to qubit coherence from the Stark tones, our experiment combines the quantum control approach to static $ZZ$ cancellation introduced here with the hardware approach of multi-path couplers~\cite{kandala2020}. The multi-path couplers already suppress the $\tilde{\nu}_{ZZ}$ compared to an equivalent direct coupler with the same effective $J$. This reduces the amplitude of the siZZle tones required to then tune to $\tilde{\nu}_{ZZ} \sim 0$, and consequently, the magnitude of the individual qubit Stark shifts (see Eq. ~\ref{eqn:zi_single}), and the impact to qubit coherence, if any. 
This discussion and the device properties are detailed in the Supplementary Information. As a reminder, we have three knobs to manipulate the $ZZ$ interaction in the device: the amplitude of the off-resonant tones, their detuning from the qubit frequencies, and the pair wise phase difference. This makes it particularly attractive for device-wide $ZZ$ cancellation on even more complex topologies. For the considered line of qubits, we choose a common Stark frequency set to 5.1 GHz, above all the qubit frequencies, leaving the individual amplitudes and phases as the free control parameters. Placing the Stark frequency above all the qubits reduces the possibility of undesired frequency collisions. We then adjust the Stark amplitude on one of the qubits to induce a Stark shift of $\sim$ 1 MHz. The amplitudes of the CW tones on the subsequent qubits are then adjusted sequentially such that it is just sufficient to tune to $\tilde{\nu}_{ZZ} \sim 0$ for every pair (i.e. $\phi_i - \phi_j \sim \pi$). We then re-calibrate the single and two qubit gates at the new dressed frequencies. We see that we can tune to $\tilde{\nu}_{ZZ} \sim 0$ with very modest Stark shifts ($\sim$ 1 MHz), which is important for reducing the impact to qubit coherence, as discussed above. We then use cross-resonance to calibrate an echo CNOT with rotary target drives, as in ~\cite{sundaresan:2020}. We emphasize that we observe no large changes in CNOT gate fidelity for all the pairs, at the different $\tilde{\nu}_{ZZ}$ levels, which highlights the need for circuit-level benchmarks such as quantum volume (QV) ~\cite{cross2019} that are sensitive to accumulated $ZZ$ errors from qubit idle times. In order to benchmark multi-qubit performance, we employ seven-qubit QV circuits and observe an improvement in the heavy output probability (HOP) from 0.5810 $\pm$ 0.0027 to 0.5996 $\pm$ 0.0023 as the average $\tilde{\nu}_{ZZ}$ is tuned from the bare value $\sim$ 40 kHz to $\sim$ 0 kHz. We employ 200 random circuits, with a mean circuit time of $\sim$ 14.1 $\mu s$, and $83$ CNOT gates on average. The improvement in the distribution of the individual circuit HOPs with $ZZ$ cancellation is also depicted in Fig. \ref{fig:HOP}. For the purpose of this demonstration, we do not employ the circuit optimization and improved readout techniques discussed in ~\cite{jurcevic:2020}. Our control knobs also enable us to systematically study the impact of $\tilde{\nu}_{ZZ}$ on circuit performance. We modulate the average $\tilde{\nu}_{ZZ}$ in the device merely by adjusting the pair-wise phase differences, and re-calibrate all the gates at every step. Fig. \ref{fig:HOP} depicts the systematic decrease in HOP with increase in average $\tilde{\nu}_{ZZ}$, and highlights why $ZZ$ cancellation will be crucial for improving the performance of superconducting quantum processors. The technique also opens up the path to more targeted studies of the impact of the $ZZ$ interaction on spectator interactions and parallel gate operation, all in a single device.

In conclusion, we demonstrate an all microwave technique - siZZle - for arbitrary control of the $ZZ$ interaction rate in coupled transmon devices. We use siZZle to demonstrate a novel high-fidelity CZ gate that could enable hardware-efficient implementations of near-term algorithms on existing fixed-frequency quantum processors. Furthermore, static $ZZ$ cancellation  with siZZle enables us to take cross-resonance past the 100 ns milestone for two-qubit gate time, with state-of-the-art fidelity. This gives us a clear path to increasing the fixed $J$ coupling in devices and also serves as a platform to explore the physics of well-controlled strong coupling interactions. Finally, combining siZZle with hardware approaches to $ZZ$ cancellation is leveraged to definitively improve multi-qubit circuit fidelity, and highlights the scalability of our technique. These results reveal quantum control with multi-color drive tones to be an attractive approach to extend the reach of fixed frequency superconducting quantum architectures. 



We note recent independent work ~\cite{mitchell:2021} reporting siZZle and a CZ gate based on the effect.

\begin{acknowledgments}
We acknowledge Malcolm Carroll, Antonio Corcoles, Pat Gumann, Micheal Gordon, Shawn Hall, Sean Hart, Muir Kumph, Jim Rozen, Maika Takita for experimental contributions and Doug McClure, Petar Jurcevic for helpful discussions. The device  bring-up,  gate calibration and  characterization work was supported by IARPA under LogiQ (contract W911NF-16-1-0114).
\end{acknowledgments}

\bibliography{sizzle}



\pagebreak
\widetext
\begin{center}
\textbf{\large Supplementary Information: Quantum crosstalk cancellation for fast entangling gates and improved multi-qubit performance}
\end{center}

\setcounter{equation}{0}
\setcounter{figure}{0}
\setcounter{table}{0}
\setcounter{page}{1}
\makeatletter
\renewcommand{\theequation}{S\arabic{equation}}
\renewcommand{\thefigure}{S\arabic{figure}}
\renewcommand{\bibnumfmt}[1]{[S#1]}
\renewcommand{\citenumfont}[1]{S#1}

\section{SiZZle theory}\label{sec:StarkZZ}

We first provide some intuition for the dual-drive Stark induced $ZZ$ effect. We consider a simple two-level model for the qubits, dressed by monochromatic drives, with the $J$ coupling introduced as a perturbation. In the absence of coupling, each qubit can be described independently by 

\begin{equation}
    H_0/h = \nu_{Q0} |1\rangle \langle 1| + \Omega_0 \cos(2\pi \nu_d t + \phi_0) (|0\rangle \langle 1| + |1\rangle \langle 0|).
\end{equation}
The off-resonant drive dresses the eigenstates and shifts the eigenvalues,
\begin{eqnarray}
|\overline{0}\rangle & \approx & |0\rangle - \frac{\Omega_0}{2\Delta_0}e^{-i\phi_0} |1\rangle, \\
|\overline{1}\rangle & \approx & |1\rangle + \frac{\Omega_0}{2\Delta_0}e^{i\phi_0} |0\rangle, \\
\overline{E}_0/h & \approx & -\frac{\Omega^2_0}{4\Delta_0}, \\
\overline{E}_1/h & \approx & \nu_{Q0} + \frac{\Omega^2_0}{4\Delta_0},
\end{eqnarray}
where $\Delta=\nu_{Q0}-\nu_d$ is the detuning. Therefore, in the two qubit basis, the dressed states are,
\begin{eqnarray}
|\overline{00}\rangle & \approx & |00\rangle - \frac{\Omega_0}{2\Delta_0}e^{-i\phi_0} |10\rangle - \frac{\Omega_1}{2\Delta_1}e^{-i\phi_1} |01\rangle + \frac{\Omega_0 \Omega_1}{4 \Delta_0 \Delta_1} e^{-i(\phi_0+\phi_1)} |11\rangle,\\
|\overline{10}\rangle & \approx & |10\rangle + \frac{\Omega_0}{2\Delta_0}e^{i\phi_0} |00\rangle - \frac{\Omega_1}{2\Delta_1}e^{-i\phi_1} |11\rangle - \frac{\Omega_0 \Omega_1}{4 \Delta_0 \Delta_1} e^{i(\phi_0-\phi_1)} |01\rangle,\\
|\overline{01}\rangle & \approx & |01\rangle - \frac{\Omega_0}{2\Delta_0}e^{-i\phi_0} |11\rangle + \frac{\Omega_1}{2\Delta_1}e^{i\phi_1} |00\rangle - \frac{\Omega_0 \Omega_1}{4 \Delta_0 \Delta_1} e^{-i(\phi_0-\phi_1)} |10\rangle,\\
|\overline{11}\rangle & \approx & |11\rangle + \frac{\Omega_0}{2\Delta_0}e^{i\phi_0} |01\rangle + \frac{\Omega_1}{2\Delta_1}e^{i\phi_1} |10\rangle + \frac{\Omega_0 \Omega_1}{4 \Delta_0 \Delta_1} e^{i(\phi_0+\phi_1)} |00\rangle.
\end{eqnarray}

The dressing of the  $|00\rangle$ and $|11\rangle$ with $|01\rangle$ and $|10\rangle$ allows exchange interactions to directly couple them, leading to a $ZZ$ interaction. We explicitly show this by calculating the energy shifts due to a $J$ coupling, $H_J/h = J (|01\rangle \langle 10| + |10\rangle \langle 01|)$ ,
\begin{eqnarray}
\langle \overline{01} | H_J/h | \overline{01} \rangle  & \approx &  -J\frac{\Omega_0 \Omega_1}{4 \Delta_0 \Delta_1} e^{-i(\phi_0-\phi_1)}, \\
\langle \overline{10} | H_J/h | \overline{10} \rangle & \approx & -J\frac{\Omega_0 \Omega_1}{4 \Delta_0 \Delta_1} e^{i(\phi_0-\phi_1)}, \\
\langle \overline{00} | H_J/h | \overline{00} \rangle & \approx & J\frac{\Omega_0 \Omega_1}{2 \Delta_0 \Delta_1} \cos(\phi_0 - \phi_1), \\
\langle \overline{11} | H_J/h | \overline{11} \rangle & \approx & J\frac{\Omega_0 \Omega_1}{2 \Delta_0 \Delta_1} \cos(\phi_0 - \phi_1).
\end{eqnarray}
From Eq. ~\ref{eqn:zz_def}, we thus obtain for the doubly dressed frame,
\begin{eqnarray}
\tilde{\nu}_{ZZ} & \approx & 2J\frac{\Omega_0 \Omega_1}{\Delta_0 \Delta_1} \cos(\phi_0 - \phi_1).
\end{eqnarray}
For transmons we perform a similar calculation and including the $|2\rangle$ state leads to the expression in Eq.~\ref{eqn:ZZ} of the main text.

More generally, we start from Eq.~\ref{eqn:mainh} of the main text,
\begin{eqnarray}
H_{\textrm{siZZle}}/h & = &  H_0/h + \sum_{i = \{0,1\}} \Omega_{i}\cos{(2\pi\nu_{d}t+\phi_i)}(\hat{a}_i^{\dagger}+\hat{a}_i) ,
\end{eqnarray} 
with amplitudes $\Omega_i$, phases $\phi_i$, and a common frequency $\nu_d$. First, we move into the frame rotating at $\nu_d$ via the unitary operator,
\begin{align}
    R_d&=e^{-i2\pi\nu_d t \left(\hat{a}_0^{\dagger}\hat{a}_0+\hat{a}_1^{\dagger}\hat{a}_1\right)}.
\end{align}
The RWA is made on the drive tones by ignoring fast rotating terms. The result is a time-independent Hamiltonian and diagonalizing followed by restoring $\nu_d$ via $R_d^\dagger$ gives the effective Hamiltonian describing the dynamics under the exchange coupling and Stark tones,
\begin{align}
    H_\text{eff}/h &= \tilde{\nu}_{IZ}IZ/4 + \tilde{\nu}_{ZI}ZI/4 + \tilde{\nu}_{ZZ}ZZ/4,
\end{align}
where $\tilde{\nu}_{ZZ} = (\tilde{E}_{00} + \tilde{E}_{11} - \tilde{E}_{01} - \tilde{E}_{10})/h$. For the case $\alpha_0 \approx \alpha_1$, $\tilde{\nu}_{ZZ}$ is given in Eq.~\ref{eqn:ZZ} of the main text and $\tilde{\nu}_{IZ}$, $\tilde{\nu}_{ZI}$ are given by,
\begin{align}
    \tilde{\nu}_{IZ}&=  (\tilde{E}_{01} - \tilde{E}_{00} + \tilde{E}_{11} - \tilde{E}_{10})/h \approx \nu_{IZ,J} + \nu_{1,s} + \frac{ J (\alpha_0+\alpha_1)  \Omega_0\Omega_1\cos(\phi_0-\phi_1)}{\Delta_{1,d} (\alpha_0 +\Delta_{0,d}) (\alpha_1 +\Delta_{1,d})},
\end{align}
\begin{align}
    \tilde{\nu}_{ZI} &=  (\tilde{E}_{10} - \tilde{E}_{00} + \tilde{E}_{11} - \tilde{E}_{01})/h \approx \nu_{ZI,J} + \nu_{0,s} + \frac{ J (\alpha_0+\alpha_1)  \Omega_0\Omega_1\cos(\phi_0-\phi_1)}{\Delta_{0,d} (\alpha_0 +\Delta_{0,d}) (\alpha_1 +\Delta_{1,d})},
\end{align}
where
\begin{align}
\tilde{\nu}_{IZ,J} &= 2\left(-\nu_1 + J^2\left(\frac{1}{\Delta_{01}} + \frac{\alpha_0+\alpha_1}{(\Delta_{01}+\alpha_0)(\Delta_{01}-\alpha_1)}\right)\right), \nonumber \\
\tilde{\nu}_{ZI,J} &= 2\left(-\nu_0 + J^2\left(-\frac{1}{\Delta_{01}} + \frac{\alpha_0+\alpha_1}{(\Delta_{01}+\alpha_0)(\Delta_{01}-\alpha_1)}\right) \right),
\end{align}
\begin{align}
\nu_{0,s} &= -\frac{\Omega_0^2\alpha_0}{\Delta_{0,d}(\alpha_0+\Delta_{0,d})}, \nonumber \\
\nu_{1,s}&= -\frac{\Omega_1^2\alpha_1}{\Delta_{1,d}(\alpha_1+\Delta_{1,d})}.
\end{align}

\section{Cross-resonance with $ZZ$ cancellation tones}

The starting Hamiltonian is given by,
\begin{eqnarray}
H/h & = & \sum_{i \in \{0,1\}} \left(\nu_i \hat{a}_{i}^{\dagger}\hat{a}_{i} + \frac{\alpha_i}{2}\hat{a}_{i}^{\dagger}\hat{a}_{i} \left(\hat{a}_{i}^{\dagger}\hat{a}_{i}-1\right)\right)  +   J (\hat{a}_0^{\dagger}+\hat{a}_0)(\hat{a}_1^{\dagger}+\hat{a}_1)\nonumber \\
&& +  \sum_{i\in\{0,1\}} \Omega_{i}\cos{(2\pi\nu_{d}t+\phi_i)}(\hat{a}_i^{\dagger}+\hat{a}_i) +   \Omega_{0,\textrm{gate}}(t)\cos{(2\pi\nu_{0,\textrm{gate}}t+\phi_{0,\textrm{gate}})}(\hat{a}_0^{\dagger}+\hat{a}_0).
\end{eqnarray}
In order to find the effective Hamiltonian describing the system including the cross-resonance tone, we first find the effective Hamiltonian describing the dynamics of the exchange coupling and Stark tones. The series of transformations are also applied to the CR drive tone $\Omega_{0,\textrm{gate}}(t)\cos{(2\pi\nu_{0,\textrm{gate}}t+\phi_{0,\textrm{gate}})}(\hat{a}_0^{\dagger}+\hat{a}_0)$ so we obtain,
\begin{align}
    H &\rightarrow H_\text{eff} + \Omega_{0,\textrm{gate}}(t)\cos{(2\pi \nu_{0,\textrm{gate}}t+\phi_{0,\textrm{gate}})} D_\textrm{CR}(t),
\end{align}
where $D_\textrm{CR}(t)$ is the transformed drive operator. We set $\nu_{0,\textrm{gate}} = \tilde{\nu}_1$ and $\phi_{0,\textrm{gate}} = 0$.
Moving into the frame rotating at $\tilde{\nu}_1$ and making the RWA gives to first-order in the cross-resonance tone amplitude, first order in $J$, second order in the Stark tone amplitudes, and assuming $\alpha_0=\alpha_1=\alpha$ for simplicity,
\begin{align}
    \tilde{\nu}_{ZX} &= \text{tr}\left(H_\text{eff,CR} \frac{ZX}{2}\right)= J\Omega_{0,\text{gate}}\left(A+B\Omega_0^2 + C\Omega_1^2\right),
\end{align}
where
\begin{align}
A &= -\frac{\alpha}{\Delta_{0,1}(\alpha+\Delta_{0,1})},
\end{align}
\begin{align}
B &= -\frac{\alpha}{4\Delta_{0,1}(\alpha+\Delta_{0,1})^2\Delta_{0,d}} + \frac{(2\alpha+\Delta_{0,1})}{8(\alpha+\Delta_{0,1})\Delta_{0,d}\Delta_{0,1}^2} - \frac{\alpha}{4(\alpha+\Delta_{0,d})(\alpha+\Delta_{0,1})\Delta_{0,1}^2} \nonumber \\
&- \frac{\alpha}{4(\alpha+\Delta_{0,1}+\Delta_{0,d})(\alpha+\Delta_{0,1})\Delta_{0,1}^2} + \frac{(2\alpha+\Delta_{0,1})}{8\Delta_{1,d}(\alpha+\Delta_{0,1})\Delta^2} + \frac{\Delta_{0,1}(\alpha+\Delta_{0,d}+\Delta_{1,d})}{8(\alpha+\Delta_{0,1})^2(2\alpha+\Delta_{0,1})(\alpha+\Delta_{0,d})\Delta_{1,d}} \nonumber \\
&+\frac{1}{16(\alpha+\Delta_{0,1})^2} \Bigg(-\frac{2}{\Delta_{0,d}} - \frac{2}{\alpha+\Delta_{0,d}} - \frac{2\alpha}{(2\alpha+\Delta_{0,1})(\alpha+\Delta_{0,d})} + \frac{6\alpha}{(2\alpha+\Delta_{0,1})(2\alpha+\Delta_{0,d})} \nonumber \\
&+  \frac{2\alpha}{(\alpha+\Delta_{0,d})(\alpha+\Delta_{0,1}+\Delta_{0,d})} +  \frac{6\alpha}{(2\alpha+\Delta_{0,1})(3\alpha+\Delta_{0,1}+\Delta_{0,d})} - \frac{10\alpha+4\Delta_{0,1}}{\Delta_{1,d}(2\alpha+\Delta_{0,1})}\Bigg),
\end{align}
and
\begin{gather}
C = \frac{\alpha}{4\Delta_{0,1}^2} \Bigg(\frac{1}{(\Delta_{0,1}-\alpha)\Delta_{0,d}} - \frac{\Delta_{0,1}}{(\alpha+\Delta_{0,1})^2(\alpha+\Delta_{0,d})} + \frac{\alpha (\alpha+3\Delta_{0,1})}{(\Delta_{0,1}-\alpha)(\alpha+\Delta_{0,1})^2\Delta_{1,d}} \nonumber \\
- \frac{\alpha (\alpha+3\Delta_{0,1})}{(\Delta_{0,1}-\alpha)(\alpha+\Delta_{0,1})^2(\alpha+\Delta_{1,d})} 
+ \frac{\Delta_{0,1}}{(\alpha+\Delta_{0,1})^2(\Delta_{1,d}-\Delta_{0,1})} + \frac{1}{(\alpha-\Delta_{0,1})(\alpha-\Delta_{0,1}+\Delta_{1,d})}\Bigg).
\end{gather}

\section{Gate Calibration: Device A}
The single qubit gates are 4$\sigma$ derivative Gaussian quadrature corrected (DRAG) pulses~\cite{chow:2010} of duration 50 ns. The CNOT gate consists of two flat-topped Gaussian pulses applied simultaneously on the control and target qubits at the target frequency, followed by Z rotations on both qubits implemented by frame changes~\cite{mckay:2017}. The target pulse envelope is given by
\begin{align*}
    \Omega(t)=\Omega_x(t)\cos(2\pi\nu_{tg} t) + (\beta \dot\Omega_x(t) + \gamma |\dot\Omega_x(t)|)\sin(2\pi\nu_{tg} t)
\end{align*}
where $\Omega_x$ is the flat-topped Gaussian pulse, $\nu_{tg}$ is the target frequency, $\beta$ and $\gamma$ are the DRAG and skew corrections respectively. The control pulse does not have DRAG or skew correction. 

\begin{figure}
    \centering
    \includegraphics[scale=0.66]{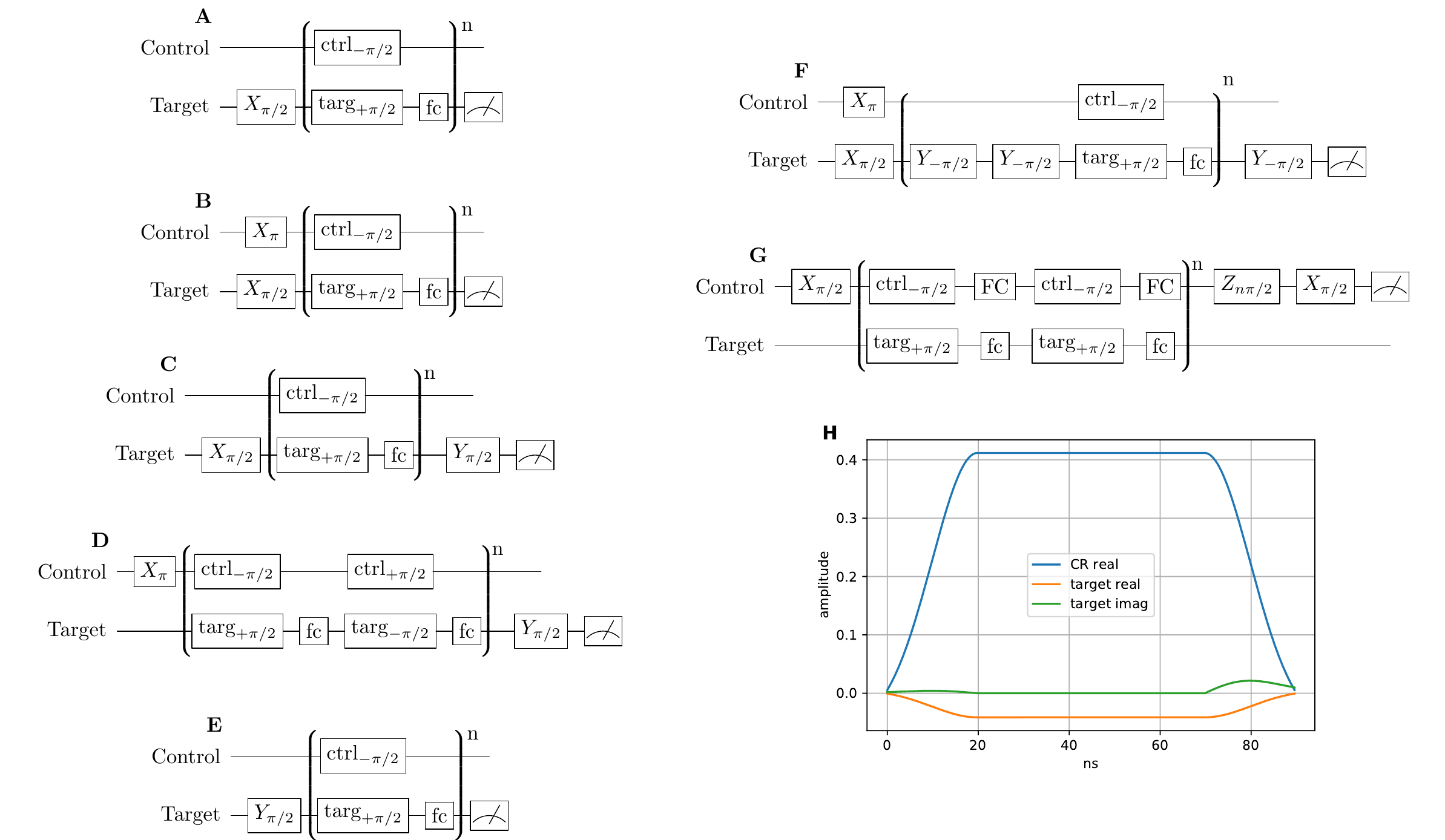}
    \caption{{\bf{Direct CNOT gate calibration sequences}} Sequences A-F are implemented simultaneously. Target and control amplitudes are updated according to the outputs of A and B. The target frame change, target drag, target skew, and control/target phase are updated according to the outputs of C-F respectively. Sequence G is used to calibrate the control frame change. The calibrated pulse envelope for the CNOT gate is shown in H. The sequence in bracket is repeated $n$ times. For A-F the target qubit population is measured, for G the control qubit population is measured. The sequence in bracket is repeated n times}
    \label{fig:calseqCX}
\end{figure}

To begin with the CNOT gate calibration, we do a rough amplitude calibration on the control pulse such that the $ZX$ rotation on the target is $\pi/2$, then we apply a pulsed version of Hamiltonian tomography~\cite{sheldon:2016} on the control pulse to align the $ZX$ interaction along the $-x$ axis. Next we turn on the target pulse and do a fine calibration by simultaneously varying the control amplitude, target amplitude, control and target phases, target drag, target skew, and target frame change to tune the gate unitary to be $\ketbra{0}{0}\otimes I + e^{-i\varphi} \ketbra{1}\otimes X$. Finally, the control frame change is calibrated to cancel $\varphi$, which brings the unitary to a CNOT gate. The fine calibration sequences are shown in Fig.~\ref{fig:calseqCX} A-F, which measures the target dynamics when the control qubit is in either $\ket{0}$ or $\ket{1}$ state. The control amplitude and target amplitude are updated according to  Fig.~\ref{fig:calseqCX} A and B, the goal to simultaneously satisfy $\theta_{ZX}+\theta_{IX}=0$ and $-\theta_{ZX}+\theta_{IX}=\pi$, where $\theta_{ZX}$ and $\theta_{IX}$ are the rotations due to cross-resonance and target pulses respectively. The target drag ($\beta$) and the gate angle are updated according to  Fig.~\ref{fig:calseqCX} D and F, these calibrations make sure the target rotation is along the $x$-axis when the control is in $\ket{1}$. When calibrating the gate angle, the control and target phases are updated together. Finally, the target skew ($\gamma$) and target frame change (fc) are calibrated according to  Fig.~\ref{fig:calseqCX} E and C, which ensures the target dynamics is identity when control is in $\ket{0}$. The control frame change (FC) is calibrated at the very end, using the sequence in  Fig.~\ref{fig:calseqCX} G. The final calibrated pulse envelope is shown in FIG.~\ref{fig:calseqCX} H, the rise and fall for the flat-topped Gaussian pulses are $2\sigma$ long where $\sigma=10$ns.

\begin{figure}
    \centering
    \includegraphics[scale=0.64]{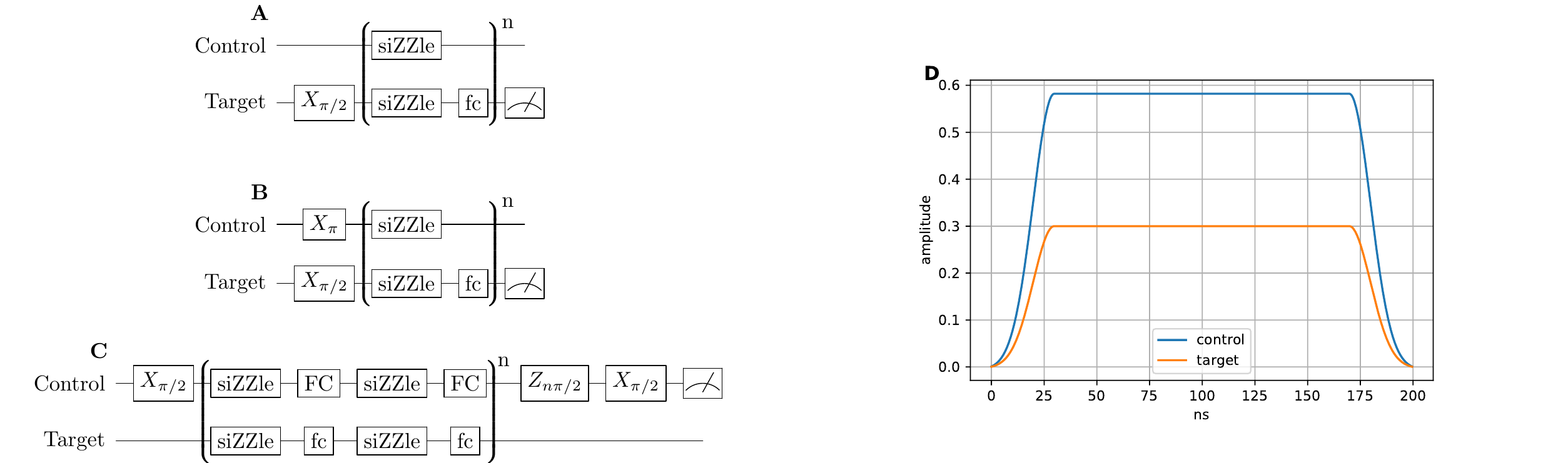}
    \caption{{\bf{Direct CZ gate calibration sequences}} Sequence A and B are implemented simultaneously, and the outputs are used to update the control amplitude and target frame change (fc). Sequence C is used to calibrate the control frame change (FC). The calibrated pulse envelope for the CZ gate is shown in D.}
    \label{fig:calseqCZ}
\end{figure}

The calibration of the CZ gate is similar to that of the CNOT gate. Here, two flat-topped Gaussian pulses are applied simultaneously to the control and target qubits at the siZZle frequency, followed by two frame changes on the control and target qubits. We fix the target amplitude and the relative phase between the two siZZle pulses, then calibrate the control amplitude and target frame change simultaneously to satisfy $\theta_{ZZ}+\theta_{IZ}=0$ and $-\theta_{ZZ}+\theta_{IZ}=\pi$. Finally we calibrate the control frame change to cancel the control Stark shift and bring the unitary to a CZ gate. The calibration sequence are shown in Fig.~\ref{fig:calseqCZ} A-C, and the final CZ pulse envelope are shown in Fig.~\ref{fig:calseqCZ} D, where the rise and fall times are $3\sigma$ with $\sigma=10$ns. Unlike for CNOT gate, drag and skew are not used in the CZ gate calibration.

\begin{figure}
    \centering
    \includegraphics[width=1.0\columnwidth]{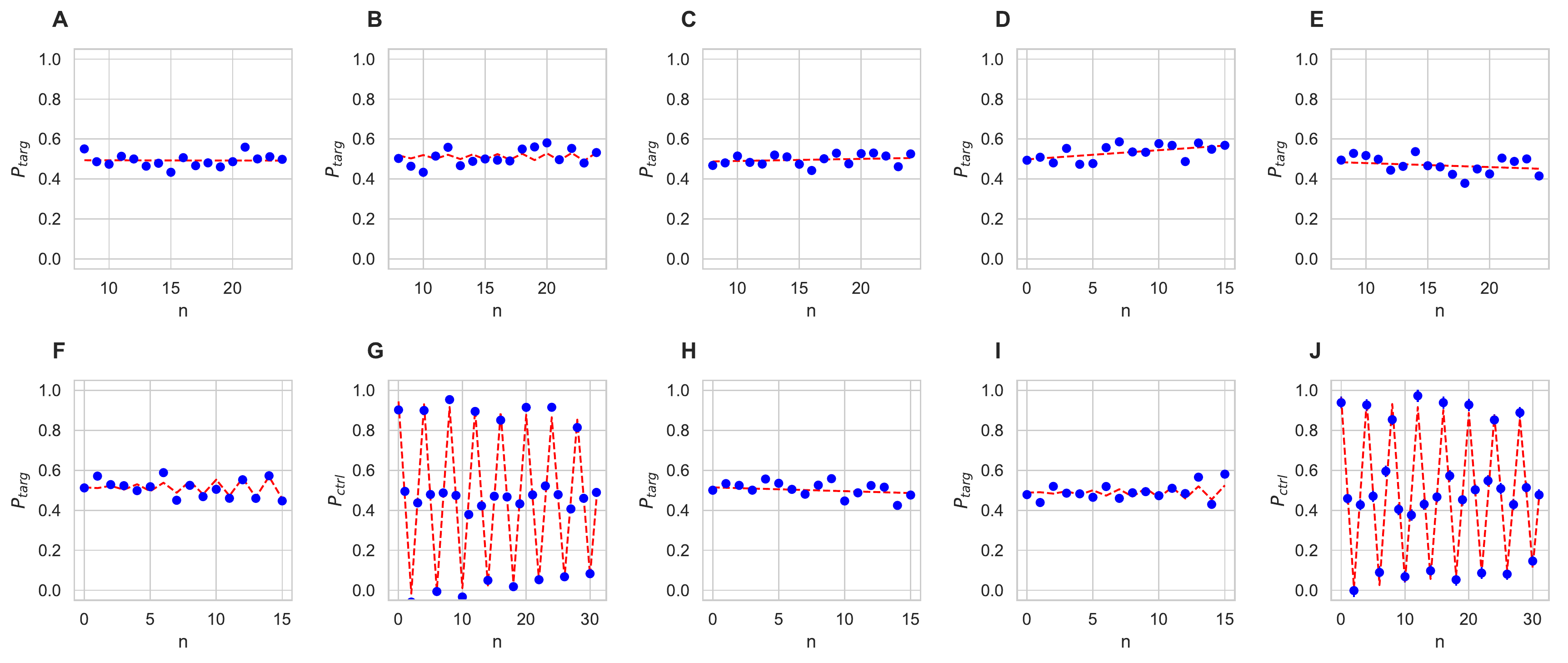}
    \caption{{\bf{Output of the calibration sequences used for CNOT and CZ gates.}} A-G correspond to the output of the CNOT calibration sequences respectively. Where as H-J correspond to the output of the CZ calibration sequences respectively. The blue points are experimental data, and the red dashed lines are the fits to experimental data. We extract the calibrated rotation angles from the fits. The y-axis in each plot corresponds to either the control or target population, and the x-axis is number of repetitions (n) shown in the calibration sequence.}
    \label{fig:caloutput}
\end{figure}

We show the calibration data for both the CNOT and CZ gates. The fine calibration routine is an iterative process~\cite{sheldon:2015}, which terminates when the absolute difference between the calibrated rotation angles and the desired rotation angles becomes less than $0.01$. In FIG~\ref{fig:caloutput} A-G. we show the converged data for sequence used in the CNOT gate calibration, and in FIG~\ref{fig:caloutput} H-J the final converged data for sequence used in the CZ gate calibration.

\begin{figure*}[!ht]
\includegraphics[width = 1\columnwidth]{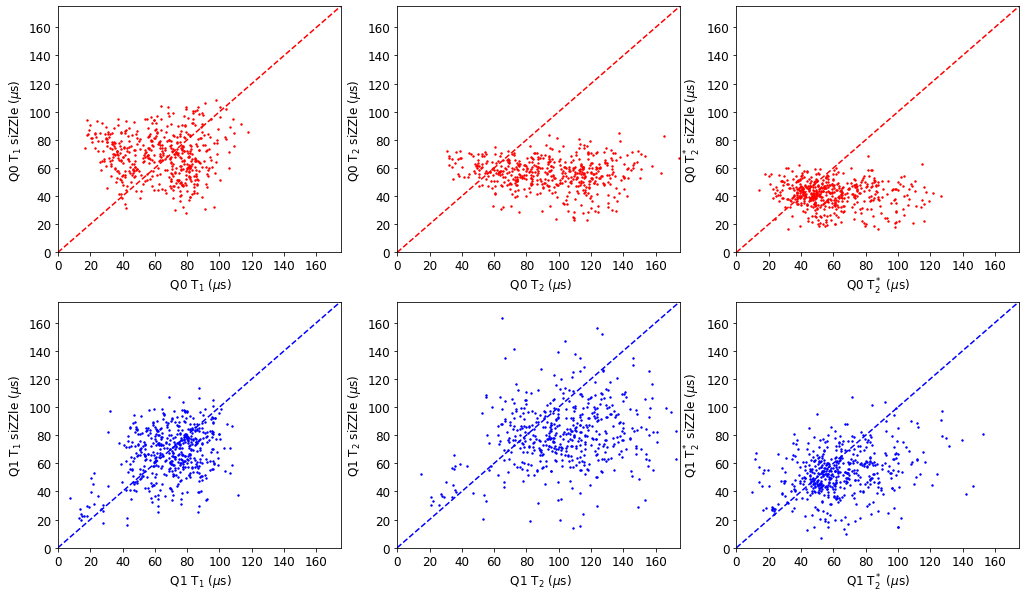}
\caption{{\bf{Device A coherence}} Scatter plots of $T_1$ (left), $T_2$ (middle), and $T_2^*$ (right) times for Q0 (top, red) and Q1 (bottom ,blue), with and without CW siZZle tones. All the measurements were interleaved and taken at 30 minute intervals. The Stark drive on Q0 for $ZZ$ cancellation is larger than Q1 at the chosen operating point, as well as the corresponding Stark shift, resulting in a clear reduction $T_2$ and $T_2^*$.
 }
\label{fig:J8_coherence}
\end{figure*}

\section{SiZZle with multi-path ZZ cancellation couplers}

As seen in the coherence data of Device A, discussed in Fig.~\ref{fig:J8_coherence}, there can be a degradation of coherence with $ZZ$ cancellation. Particularly, at large $J$ couplings with standard couplers, one requires large siZZle amplitudes $\Omega$ to achieve $ZZ$ cancellation. Since the qubit Stark shifts are proportional to $\Omega^2$, this makes the qubits more susceptible to amplitude noise, and consequently can lead to additional dephasing. In this section, we numerically show that using multi-path couplers, for the same effective $J$, one can achieve full $ZZ$ cancellation at smaller siZZle amplitudes due to requiring smaller qubit Stark shifts. We start with the following form of the Hamiltonian with a direct qubit coupling and an additional coupling path via a bus resonator.

\begin{eqnarray}
H/h & = & \sum_{i = \{0,1\}} \left(\nu_i \hat{a}_{i}^{\dagger}\hat{a}_{i} + \frac{\alpha_i}{2}\hat{a}_{i}^{\dagger}\hat{a}_{i} \left(\hat{a}_{i}^{\dagger}\hat{a}_{i}-1\right)\right) +  J (\hat{a}_0^{\dagger}+\hat{a}_0)(\hat{a}_1^{\dagger}+\hat{a}_1) +  \sum_{j} \nu_j \hat{b}_{j}^{\dagger}\hat{b}_{j} \nonumber \\
&& + \sum_{i=\{0,1\},j} g_{i,j}(\hat{a}_i^{\dagger}+\hat{a}_i)(\hat{b}_j^{\dagger}+\hat{b}_j) + \sum_{i = \{0,1\}} \Omega_{i}\cos{(2\pi\nu_{d}t+\phi_i)}(\hat{a}_i^{\dagger}+\hat{a}_i) .\label{eqn:mainh_bus}
\end{eqnarray} 
Most terms here have already been defined in Eq.~\ref{eqn:mainh} of the main text. The additional terms arise from the bus coupling, where $g_{i,j}$ is the coupling from qubit $i$ to the $j$'th harmonic mode of the bus at frequency $\nu_j$. For simplicity, we drop the counter-rotating terms and consider a single bus mode. Eq. ~\ref{eqn:mainh_bus} is then transformed into a time independent form by moving into a frame rotating at the drive frequency $\nu_d$ via the rotation operator $\hat{R}/h=e^{-i2\pi\nu_{d}t(\hat{a}_0^{\dagger}\hat{a}_0+\hat{a}_1^{\dagger}\hat{a}_1+\hat{b}^{\dagger}\hat{b})}$ and applying the RWA. One can then obtain the Stark shifts $\tilde{\nu}_{ZI}, \tilde{\nu}_{IZ}$ and the $ZZ$ interaction $\tilde{\nu}_{ZZ}$ by diagonalizing the time independent Hamiltonian. 

\begin{figure*}[h!]
\includegraphics[width = 0.45\columnwidth]{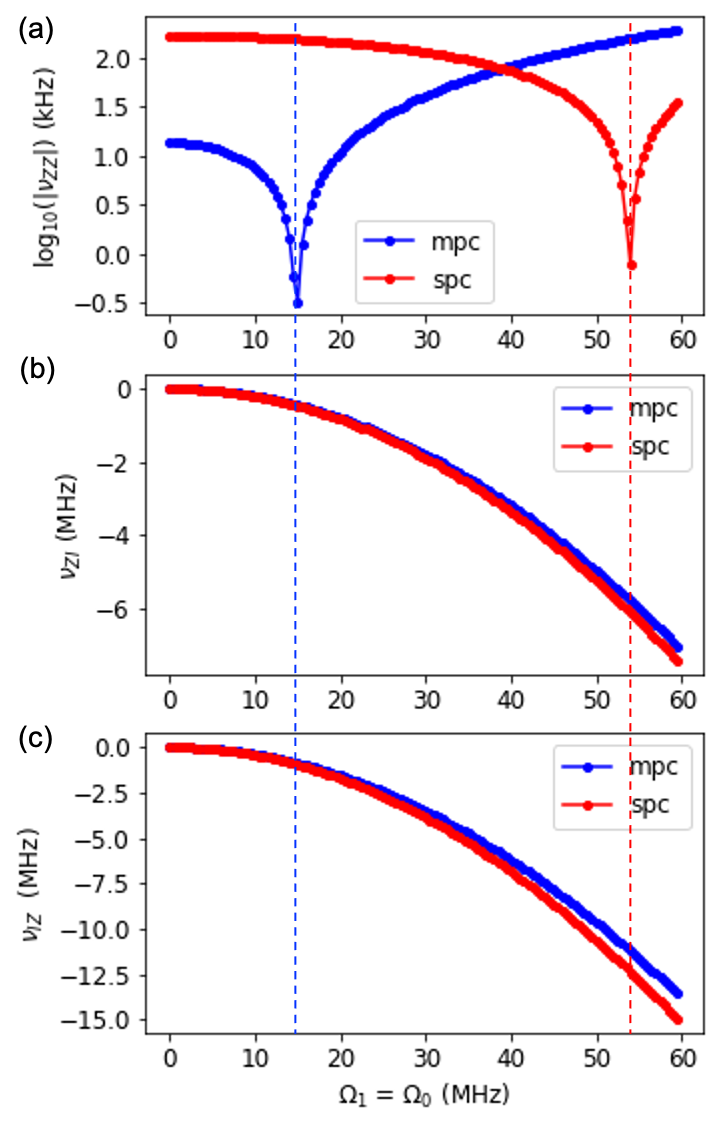}
\caption{{\bf{SiZZle with multi-path $ZZ$ cancellation couplers}} Numerical simulations of the Stark tone amplitude dependence of (a) $\tilde{\nu}_{ZZ}$, (b) $\tilde{\nu}_{ZI}$ and (c) $\tilde{\nu}_{IZ}$ for multi-path coupler (blue) and a single-path coupler (red) with the same $J_{\textrm{eff}}=3.2$ MHz, defined in the text. 
We consider the following parameters for the mpc device: $\nu_0 = 4.85$ GHz, $\nu_1 = 4.95$ GHz, $\alpha_0=\alpha_1=-290$ MHz, $g_0=g_1=135$ MHz, $J = 10.6$ MHz, $\nu_{\textrm{bus}}=6.35 $ GHz, $\nu_d=5.1$ GHz and $\Omega_0=\Omega_1$. The blue (red) dotted line represents the operating point for ZZ cancellation for the mpc (spc) device. For the mpc, the $ZZ$ cancellation is achieved at smaller Stark tone amplitudes, and the smaller $\tilde{\nu}_{ZI}$ and $\tilde{\nu}_{IZ}$ are less sensitive to amplitude noise on the Stark tones.
}
\label{fig:MapleSizzle}
\end{figure*}

We consider the following parameters, that are similar to pairs on device B: $\nu_0 = 4.85$ GHz, $\nu_1 = 4.95$ GHz, $\alpha_0=\alpha_1=-290$ MHz, $g_0=g_1=135$ MHz, $J = 10.6$ MHz, $\nu_{\textrm{bus}}=6.35 $ GHz, $\nu_d=5.1$ GHz and $\Omega_0=\Omega_1$. From the low amplitude dependence of $\tilde{\nu}_{ZZ}$, we estimate an effective $J$ coupling for the multi-path coupler(mpc) using the form of Eq.~\ref{eqn:ZZ} to be $J_{\textrm{eff}} = 3.28$ MHz. We then compare the Stark tone amplitude dependence of $\tilde{\nu}_{ZI}, \tilde{\nu}_{IZ}$ and $\tilde{\nu}_{ZZ}$ for this mpc device with a single path coupler (spc) of the same $J_{\textrm{eff}}$. This is depicted in Fig.~\ref{fig:MapleSizzle}. For the mpc, while $ZZ$ cancellation only requires Stark tone amplitudes $\sim 15$ MHz, the spc requires amplitudes $\sim 55$ MHz, seen in Fig.~\ref{fig:MapleSizzle}a. Consequently, the qubit Stark shifts are much smaller at $ZZ$ cancellation for the mpc device, seen in Fig.~\ref{fig:MapleSizzle}b and c, thereby reducing the sensitivity to Stark tone amplitude noise.

\section{Device B: Seven-qubit device with multi-path couplers for $ZZ$ cancellation}

\begin{figure*}[h!]
\includegraphics[width = 1\columnwidth]{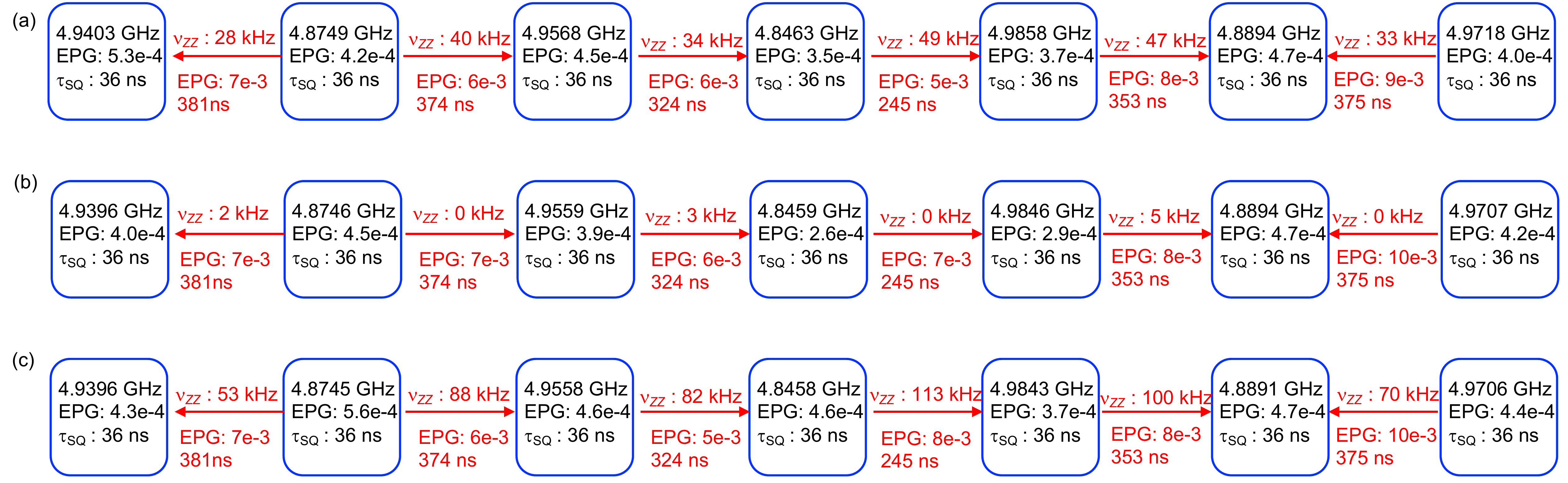}
\caption{{\bf{Device B sub-system metrics}} Qubit frequencies, single and two-qubit gate times and their respective error rates, and the strength of the pair wise static $ZZ$ interaction $\tilde{\nu}_{ZZ}$ for (a) the native device without siZZle tones (b) with siZZle tones at $\nu_d = 5.1$ GHz, and pair-wise phases tuned to $ZZ$ cancellation $\phi \sim \pi$. (c) with siZZle tones at $\nu_d = 5.1$ GHz, and pair-wise phases tuned to $ZZ$ amplification $\phi \sim 0$. All gate errors are estimated by randomized benchmarking. The arrows represent the direction of the CNOT gates employed in the QV circuits discussed in the main text, and the reported error per gates (EPG) represent the upper bound obtained from the error per Clifford (EPC/1.5). The CNOT gates are composed of two cross-resonance pulses and two finite-time single qubit pulses, and the gate times are optimized for operation in the absence of siZZle. The single qubit EPG's represent the errors for simultaneous single qubit operation. 
 }
\label{fig:F609_device}
\end{figure*}

\begin{figure*}[h!]
\includegraphics[width = 1\columnwidth]{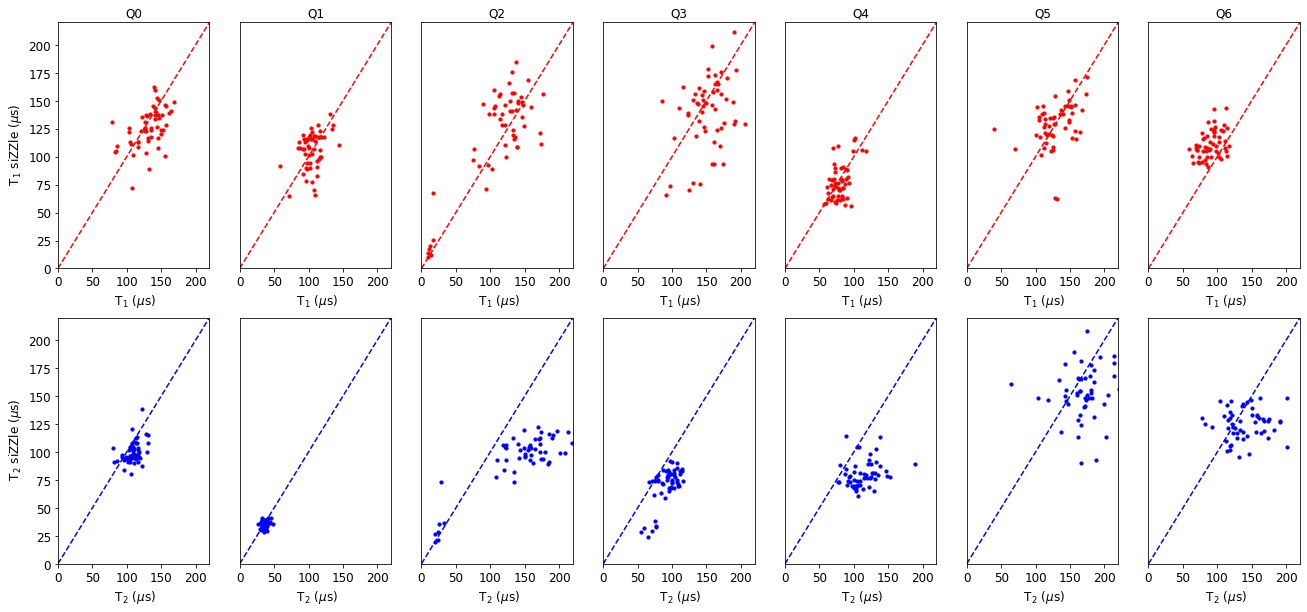}
\caption{{\bf{Device B coherence}} Scatter plots of $T_1$ (top, red) and $T_2$ (bottom, blue) times of the 7 qubits, with and without siZZle tones. All the measurements were interleaved and taken at 30 minute intervals.
}
\label{fig:F609_coherence}
\end{figure*}

In this section, we detail device B from the main text. The seven qubits represent a sub section of a larger lattice of 27 qubits in the heavy-hex architecture. In order to reduce the static $ZZ$ interaction compared to standard single path couplers, the device employs coupling elements composed of a direct capacitive coupler and a $\lambda/4$ bus resonator~\cite{kandala2020}. The bus resonator frequencies are in the range 6.35-6.55 GHz. For $ZZ$ cancellation, a common frequency $\nu_d = 5.1$ GHz was chosen for the CW tones, above all the qubit transitions. Most of the qubit parameters and gate fidelities are detailed in Fig. ~\ref{fig:F609_device}. The qubit anharmonicities are in the range -288 to -295 MHz, and the average readout fidelity is 98.2 $\%$. As seen in Fig.~\ref{fig:F609_device}, the qubit frequencies are shifted by at most 1.2 MHz, by the CW tones for full $ZZ$ cancellation. This helps retain good coherence times for the device, even after $ZZ$ cancellation, depicted in  Fig. ~\ref{fig:F609_coherence}. Some of the qubits show a modest decrease in $T_2$, while the $T_1$ times are within typical fluctuations. From the single drive Stark shifts of the qubits, we estimate the amplitude of the CW tones driving Q0/1/2/3/4/5/6 for $ZZ$ cancellation to be 17.6/16.8/20.4/19.5/21.1/13.0/21.3 MHz respectively.

\end{document}